\documentclass{aa}  
\usepackage{graphicx}
\usepackage{txfonts}
\begin{document}
\newcommand{\mearth}{M_\oplus}
   \title{Constraints on resonant--trapping for two planets embedded in a
protoplanetary disc}

   \author{A. Pierens
          \and
          R.P Nelson
          }

   \offprints{A. Pierens}

   \institute{Astronomy Unit, Queen Mary, University of London, Mile End Rd, London, E1 4NS, UK\\
     \email{a.pierens@qmul.ac.uk}       
             }

   \date{Received September 15, 1996; accepted March 16, 1997}

% \abstract{}{}{}{}{} 
% 5 {} token are mandatory
 
  \abstract
   { A number of extrasolar planet systems contain pairs of
    Jupiter--like planets in mean motion resonances. As yet there
    are no known resonant systems which consist of a 
    giant planet and a significantly lower--mass body. }
   {We investigate the evolution of two-planet systems embedded in a
   protoplanetary disc, which are composed of a 
   Jupiter-mass planet plus another body
   located further out in the disc. The aim is to examine how 
   the long--term evolution 
   of such a system depends on the mass of the outer planet.}
   {We have performed 2D numerical simulations using a grid-based
   hydrodynamics code. The  
   planets can interact with each other and with the disc in which
   they are embedded. We consider outermost planets with masses
  ranging from $10\; \mearth$ to $1\; M_J$. Combining the
  results of these calculations and analytical estimates, we  also examine the case of outermost bodies with
  masses $< \; 10 \; \mearth$.}
   {Differential migration of the planets due to 
   disc torques leads to different
   evolution outcomes depending on the mass of the outer protoplanet. 
   For planets with mass $\lesssim 3.5 \; \mearth$ the type II
   migration rate of the giant exceeds the type I
   migration rate of the outer body, resulting in divergent migration.
   Outer bodies with masses in the range $3.5 < m_o \le 20\; \mearth$ become
   trapped at the edge of the gap formed by the giant planet, because 
   of corotation torques. Higher mass planets
   are captured into resonance with the inner planet. 
   If  $30 \le m_o \le 40 \; \mearth$ or $m_o=1\;M_J$,
   then the 2:1 resonance is established.
   If $80 \le m_o \le 100 \; \mearth$,
   the 3:2 resonance is favoured. \\
   Simulations of gas-accreting protoplanets of mass 
   $m_o \ge 20 \; \mearth$, trapped initially at the edge of the gap,
   or in the 2:1 resonance, also result in eventual
   capture in the 3:2 resonance as the planet mass grows to become
   close to the Saturnian value.}
   {Our results suggest that there is a theoretical lower limit to
    the mass of an outer planet that can be captured into resonance
    with an inner Jovian planet, which is relevant to observations
    of extrasolar multiplanet systems.
    Furthermore, capture of a 
    Saturn-like planet into the 3:2 resonance with a Jupiter-like
    planet is a very robust outcome of simulations, independent
    of initial conditions. This result is relevant to recent scenarios
    of early Solar System evolution which require Saturn to have
    existed interior to the 2:1 resonance with Jupiter prior to
    the onset of the Late Heavy Bombardment.}

   \maketitle
%
%________________________________________________________________

\section{Introduction}

To date, about 25 extrasolar multiplanet systems have been
discovered (e.g.  {\tt http://exoplanet.eu/}). 
Interestingly, at least 5 of them contain two planets
trapped in a low-order mean motion resonance. For example, the planets
 in GJ 876, HD 128311, HD 82943, HD 73526 appear to be in 2:1
 resonance while two of the four planets orbiting in the 55 Cnc
 system are in 3:1 resonance. \\

The  existence of resonant systems can be understood by a model in
which two planets undergo differential migration, due to their interaction
with a protoplanetary disc, and become locked in the
resonance as their orbits converge. This scenario has been explored
by numerous
hydrodynamic simulations of  two
embedded giant  planets interacting with each other and undergoing type
II migration (Snellgrove et
al. 2001; Papaloizou 2003; Kley et al. 2004; Kley et
al. 2005), and also by N-body simulations with prescribed forces
designed to mimic disc torques (Snellgrove et al. 2001; Lee \& Peale 2002;
Nelson \& Papaloizou 2002). 
The evolution of such a system generally proceeds as
follows. Because giant planets are able to
open a gap in the disc, there is
a tendency for the gaseous material between them to be cleared, leading
ultimately to a two-planet system orbiting within a common gap. From
this time onward, the migration of the system is such that the planets
are on converging orbits. The  
final outcome is generally found to be
trapping into 2:1 resonance, which corresponds to the most common
resonance observed in extrasolar systems.\\
Masset \& Snellgrove (2001) investigated the particular case where the
inner planet has a Jupiter mass but where the mass of the outer
planet is characteristic of that of Saturn.  In that case, Saturn 
tends to undergo a fast runaway migration  (Masset \& Papaloizou
2003) and rapidly catches up with Jupiter, which drifts inward 
much more slowly. Because their orbits converge quickly, Saturn
passes through the 2:1 resonance with Jupiter but becomes 
captured into the 3:2 resonance.
These results were confirmed by Morbidelli \& Crida
(2007) who examined the same problem, but considered a larger range of
initial parameters using a code that describes more accurately the
global viscous evolution of the disc. \\
So far, there have been few studies focused on the interaction between a 
Jupiter-mass planet and a protoplanetary core with significantly lower mass
(e.g. $\sim 10\; \mearth$).
As a result of disc torques, low-mass planets are known to undergo
rapid inward migration (so--called type I migration) 
(Ward 1997; Tanaka et al. 2002).
For typical disc parameters, the type I migration 
time scale of a protoplanet with mass
$10\;\mearth$ is a few $\times 10^4$ years, 
which is significantly shorter than the
type II migration time scale (usually quoted as being $\sim 10^5$ years).
Therefore, a sufficiently massive body undergoing type I
migration can eventually become captured into resonance with a giant
planet located inside its orbit (Hahn \& Ward 1996). 
Using N-body simulations, Thommes
(2005) recently focused on the evolution of such a two-planet system.
 He showed that indeed, bodies of Earth to Neptune-mass size have a high
 likelihood  of being captured in the exterior mean motion resonances
 of the giant. In this work, the disc torques are modelled as prescribed
 forces using classical analytical expressions for the migration and
 eccentricity damping rates (Papaloizou \& Larwood 2000). Although
 the effects of Lindblad torques are well captured by such an
 approach, the corotation torques acting on the low-mass planet are
 neglected. However, as the latter approaches the edge of the
 gap opened by Jupiter, where the disc surface density gradient is
 large and positive, corotation torques can become dominant and 
 significantly change the dynamics of the system.
 As shown by Masset et al. (2006), type I migration 
 can eventually be halted in a region of the disc with a strongly positive
 surface density gradient. Migration--halting arises once the
 protoplanet reaches a fixed point where 
 the corotation torques, which are positive near the gap edge,
 exactly counterbalance the negative Lindblad torques. In this
 paper we examine how this affects the capture into resonance
 of low mass outer planets by an interior giant planet. \\

We present the results of hydrodynamical simulations of
two-planet systems  composed of a Jupiter-mass planet plus an
outermost body whose mass can vary from 
10 $\mearth$ to 1 $M_J$. We also consider the case of
  outermost bodies with masses $< \; 10 \; \mearth$ using the results of
  these simulations and analytical estimates. The aim of this work
is to investigate how the evolution of such a system depends on the
mass of the outer planet. In particular, we examine the issue of
whether or not low-mass planets can become captured into resonance
with an interior giant planet as a result of convergent migration. 
For the particular disc
model we consider, the results of these simulations suggest that
the final fate of the outermost body can be threefold: 
(i) protoplanets with masses $\lesssim 3.5 \; \mearth$ undergo type I
migration at a slower rate than the type II migration experienced
by the giant planet, so planetary configurations of this type
experience divergent migration; 
(ii) protoplanets with masses of
 $3.5 < m_o\le 20\;\mearth$ become trapped at the edge of the gap formed by
 the inner giant due to corotation torques, where they remain over
long time scales; (iii)
 higher mass bodies become captured into
 resonance with Jupiter. For planet masses in the range 
$30 \le m_o \le 40 \; \mearth$, or for $m_o= 1 \; M_J$, capture
is into the 2:1 resonance. For planet masses in the range $80 \le m_o \le
100 \; \mearth$, capture is into the 3:2 resonance. 
 Interestingly, we find that trapping into 3:2 resonance is robust with
 respect to initial conditions of the outermost body, provided the
 disc is sufficiently massive. 
 Indeed, simulations of a protoplanetary core initially trapped at the edge of
 the giant planet's gap, or in the 2:1 resonance, and accreting gas 
 until its mass becomes
 characteristic of that of Saturn,
 resulted in eventual capture into the 3:2 resonance.  \\

This paper is organized as follows. In Section 2 we describe the
hydrodynamical model. In Section 3, we present the results of our
simulations. In Section 4, we discuss out results within the context
of the early history of the outer Solar System. We finally summarize 
 and draw our conclusions in Section 5.

\section{The hydrodynamical model}
\subsection{Numerical method}
In this paper, we adopt a 2D disc model for which all the physical quantities
are vertically averaged. We work in a non-rotating frame,
  and adopt cylindrical polar coordinates
$(r,\phi)$ with the origin located at the position of the central
star. Indirect terms resulting from the fact that this frame in
non-intertial are incorporated in the equations governing the disc evolution (Nelson et al. 2000).
 These are solved  using a hydrocode called GENESIS for which a full description can be found,
for example, in De Val-Borro et al. (2006). The evolution of the
planetary orbits is computed using a fifth-order Runge-Kutta integrator
(Press et al. 1992). Here, all the planets and the disc can interact gravitationally
with each other.  The gravitational potential experienced by a
  planet from the disc is given by:
\begin{equation}
{\Phi}_{d}=-G \int_S\frac{\Sigma({\bf r}') d{\bf r}'}{\sqrt{r'^2+r_p^2-
2r'r_p\cos(\phi'-\phi_p)+\epsilon^2}}
\label{eq:pot}
\end{equation}
where $\Sigma$ is the disc surface density and where  the integral is
performed over the  surface of the disc located outside the Hill sphere of the planet. $r_p$ and $\phi_p$ are respectively
the radial and azimutal coordinates of the planet. $\epsilon$ is
smoothing length  which is set to $\epsilon=0.66\; H$, where $H$ is
the disc scale height at the planet orbit. We further note that we exclude
the material contained in the planet Hill sphere when calculating this
gravitational potential, since this material is gravitationally bound
to the planet. \\
The computational units we adopt are such that the mass of the central star
$M_\star=1$ corresponds to one Solar mass, the gravitational constant
is $G=1$ and the radius $r=1$ in the computational domain
corresponds to 5 AU. In the following,
time is measured in units of the orbital period at $r=1$.\\
In the calculations presented here, we use $N_r=256$ radial grid cells
uniformly distributed between $r_{in}=0.25$ and $r_{out}=5$ and
$N_\phi=380$ azimuthal grid cells. 

\subsection{Initial conditions}
\label{sec:init}

In the disc model employed for all the  runs presented here, the disc aspect ratio
is constant and equal to $H/r=0.05$.  We also adopt a locally isothermal
  equation of state for which the vertically
  integrated pressure $P$ is given by:
\begin{equation}
P=c_s^2\Sigma.
\end{equation}
In this equation, $c_s$ is the local isothermal sound speed and
can be written as:
\begin{equation}
c_s=\frac{H}{r}v_K,
\end{equation}
where $v_K=\sqrt{GM_\star/r}$ is the local Keplerian velocity. Using a
locally isothermal equation of state implies that we do not evolve an
energy equation in the work presented here.  The initial disc surface density
profile is chosen to be $\Sigma(r)=\Sigma_0\;r^{-1/2}$ where
$\Sigma_0=6\times 10^{-4}$ in dimensionless units. This gives a total
initial disc mass within the computational domain
$M_d\sim 2\times 10^{-2}\;
M_\odot$. Furthermore, in order
to model the anomalous viscous stress presumed to originate
because of turbulence in the disc, we use the standard `alpha'
prescription for the disc viscosity $\nu= \alpha c_s H$ (Shakura
\& Sunyaev 1973), where $c_s$
is the isothermal sound speed. In this work we set
$\alpha=10^{-3}$.\\

The inner and outermost planets initially evolve on 
circular orbits at $a_J=1$ and
$a_o=2.2$ respectively. The mass of the inner planet is 
$m_J=$ 1 $M_J$ and we consider
outer planets with masses  $m_o=$ 10, 20, 30, 40, 80, 
100 and 300 $\mearth$.
In order to give the inner planet sufficient time to open a gap,
we proceed in two steps:\\
i) we first consider a 30 $\mearth$ protoplanet held at $r=1$ on a
fixed circular orbit. This body is allowed to accrete gas from
the disc on a dynamical time scale until its mass reaches 1 $M_J$. Such
an approach is fairly consistent with evolutionary models of 
gas giant planets forming in
protoplanetary discs which suggest that rapid gas accretion occurs once the
planet mass exceeds 30--40 $\mearth$ (Papaloizou \& Nelson 2005).\\
ii) we restart this model, but with an additional outer body located at
$r=2.2$. If the mass of latter exceeds 30 $\mearth$, we adopt the
procedure described in i) above in
order to obtain the relevant value for $m_o$. We
then release both planets and let them evolve under the action of
disc torques. 
\subsection{Boundary conditions}
At the inner edge of the computational domain, we model the accretion
onto the central star by setting the radial velocity in the innermost
cells to $v_r(r_{in})=\beta \; v_{visc}(r_{in})$, where $v_{visc}(r_{in})=-3\nu/2r_{in}$
is the typical inward drift velocity for a steady--state accretion disc,
and $\beta$ is a free parameter. Contrary to the standard
open boundary conditions which are commonly used in hydrodynamical
simulations, these boundary conditions prevent the inner disc from emptying
too quickly. In order to find a suitable value for $\beta$, we 
performed a series of test simulations of an embedded Jupiter-mass
planet, and varied the value for $\beta$ . We then compared the
final state of these calculations to the results of simulations
performed by Crida et al. (2007), in which the global disc evolution is
more accurately modelled . Although the value for $\beta$ might be
problem-dependant, choosing $\beta=5$ gives good agreement with the
results obtained by Crida et al. (2007) for our particular setup.
We therefore adopt this value in the
simulations presented here.  We also use linear viscosity  between
the inner boundary and $r=0.45$ to prevent wave reflection  (Stone
\& Norman 1992).\\
In order to avoid any wave reflection at the outer edge of the
computational domain, we employ a wave killing zone in the vicinity of
the outer boundary, in which we solve the following equation at each
time step (De Val-Borro et al. 2006):
\begin{equation}
\frac{dX}{dt}=-\frac{X-X_0}{\tau}R(r)
\end{equation}
In the previous equation, $X$ represents either the surface density or
one of the velocity components and $X_0$ is the initial
value of the correponding quantity. $\tau$ is the orbital period at
the outer boundary, and $R(r)$ is a quadratic function which varies
from $R(4)=0$ to $R(5)=1$.  At the outer edge of the computational
  domain, the radial velocity is set to zero, preventing thereby
  inflow/outflow of matter there.
\begin{table}
\caption{The first column gives the run label, the second column gives
  the mass $m_o$ of the outer planet and the third column gives the
  state of the system at the end of the simulation.}             
\label{table1}      
\centering          
\begin{tabular}{c c c}     % 7 columns 
\hline\hline       
                      % To combine 4 columns into a single one 
Model & $m_o\; (\mearth$) & Result \\ 
\hline                    
 R1 & 10 & Trapping at the edge of gap\\  
 R2 & 20 & Trapping at the edge of gap\\ 
 R3 & 30 & 2:1 resonance\\ 
 R4 & 40 & 2:1 resonance\\ 
 R5 & 80 & 3:2 resonance\\ 
 R6 & 100 & 3:2 resonance\\ 
 R7 & 300 & 2:1 resonance\\ 
\hline                  
\end{tabular}
\end{table}

\begin{figure*}
   \begin{center}
   \includegraphics[width=0.75\columnwidth]{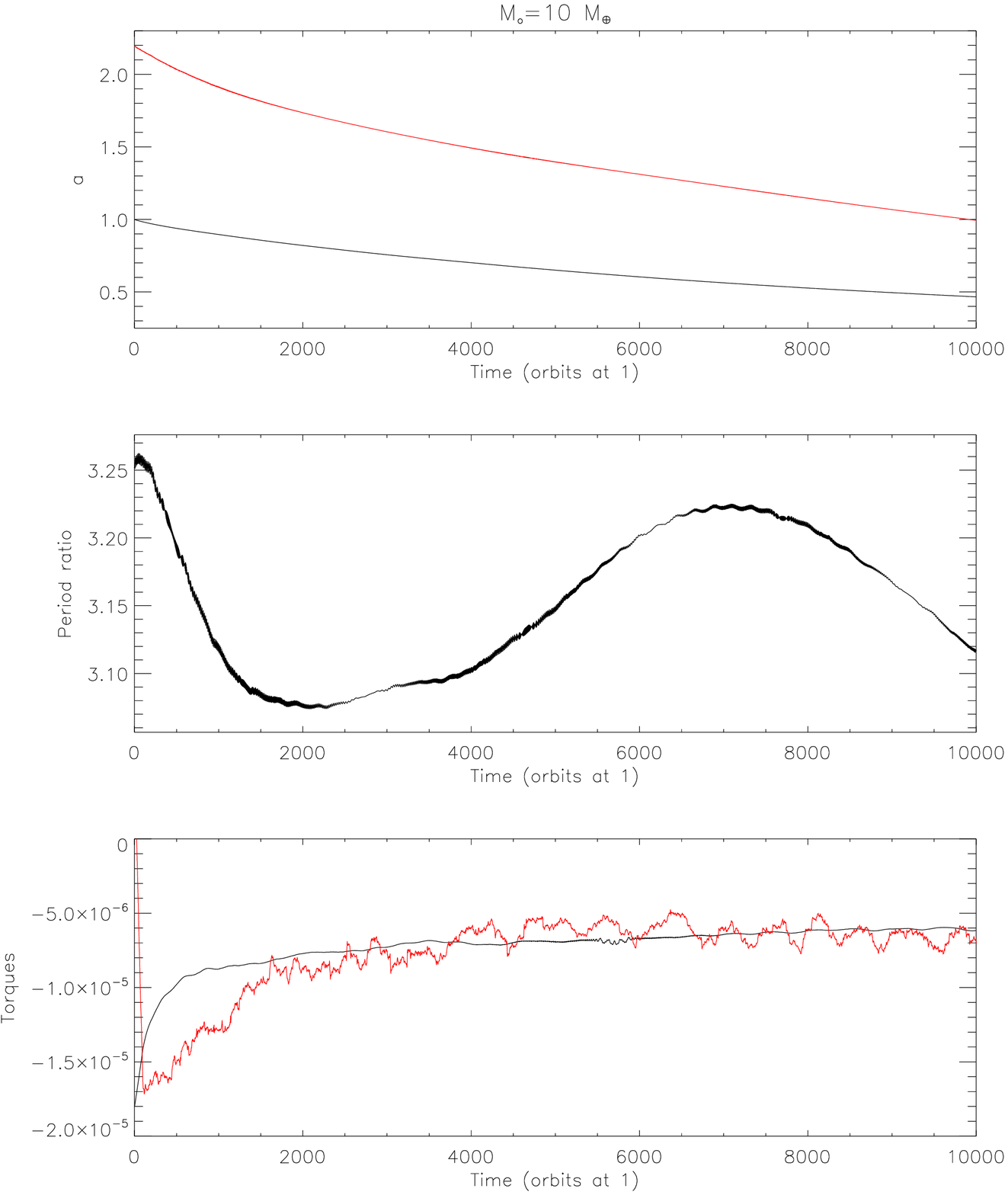}
    \includegraphics[width=0.75\columnwidth]{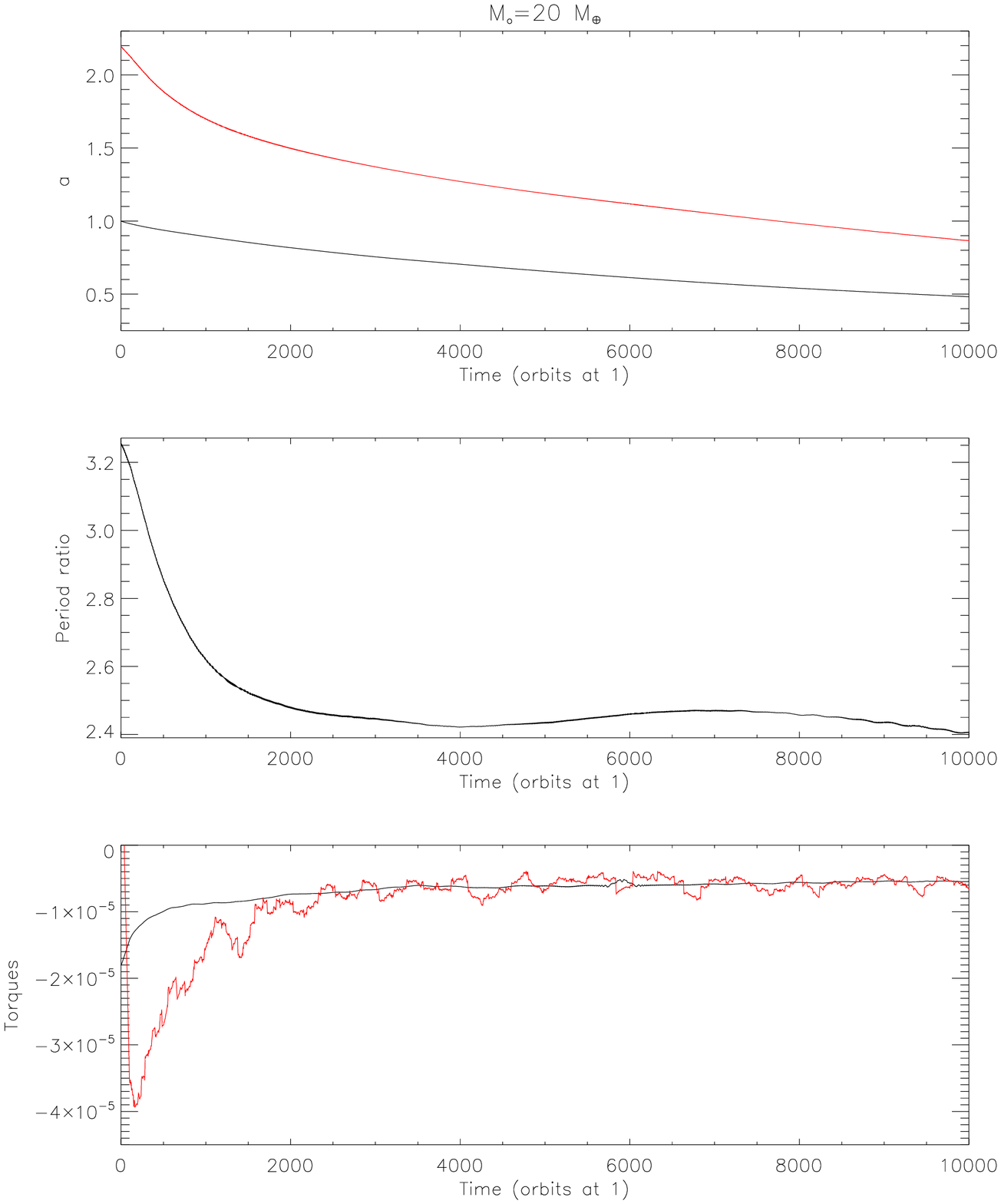}
      \caption{This figure shows the evolution of the system for
      models $R1$ ({\it left panel}) and $R2$ ({\it right panel}). In
      both cases, the giant planet is represented by black line and the outer
      planet is represented by red line. {\it Top:}
      evolution of the semi-major axes for both planets. {\it Middle:} evolution of the period
      ratio $p=(a_o/a_J)^{1.5}$. {\it Bottom:} evolution of the disc torques acting on
      the planets.}
         \label{r1r2}
     \end{center}
   \end{figure*}

\begin{figure*}
   \begin{center}
   \includegraphics[width=0.8\columnwidth]{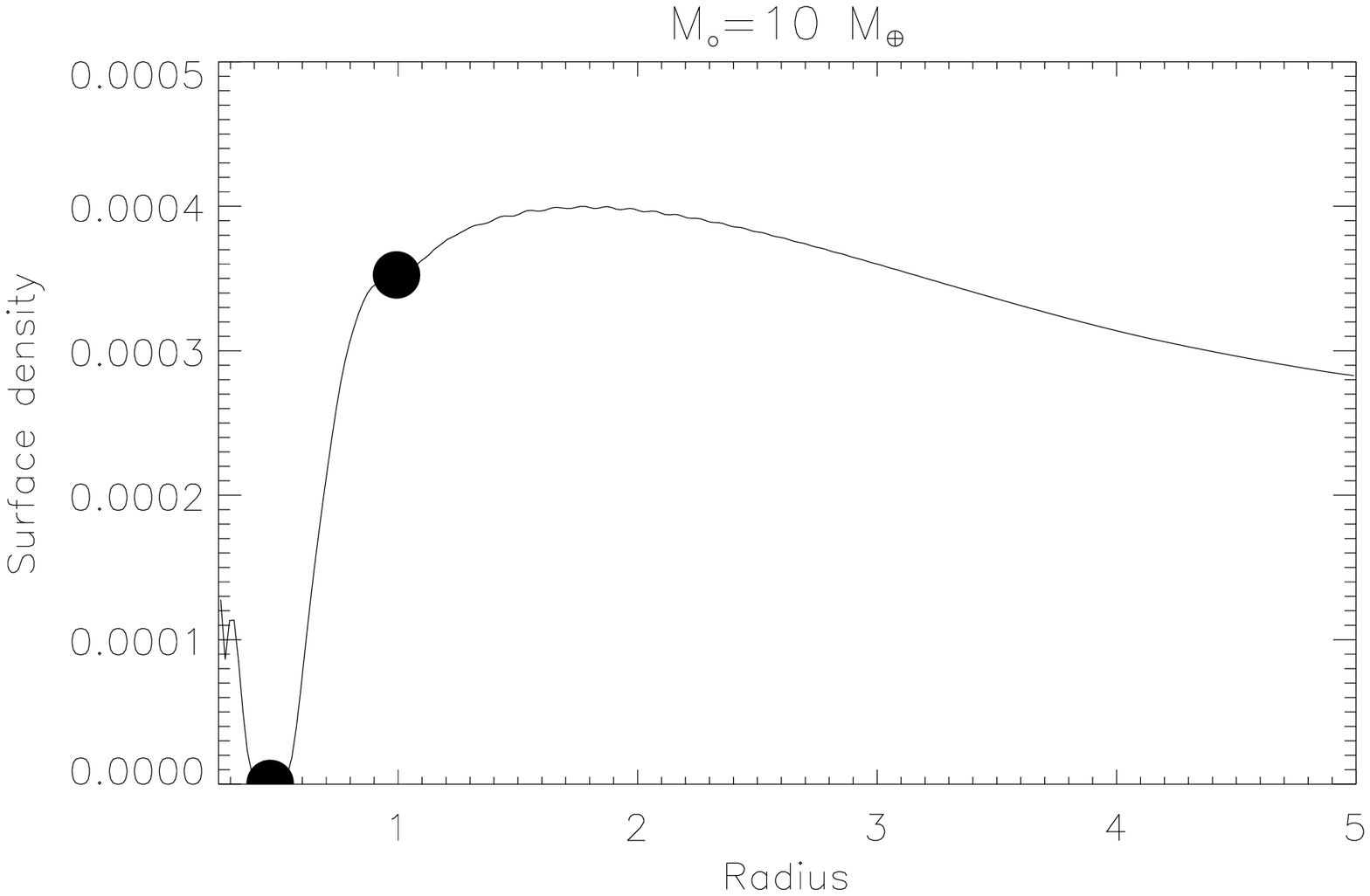}
    \includegraphics[width=0.8\columnwidth]{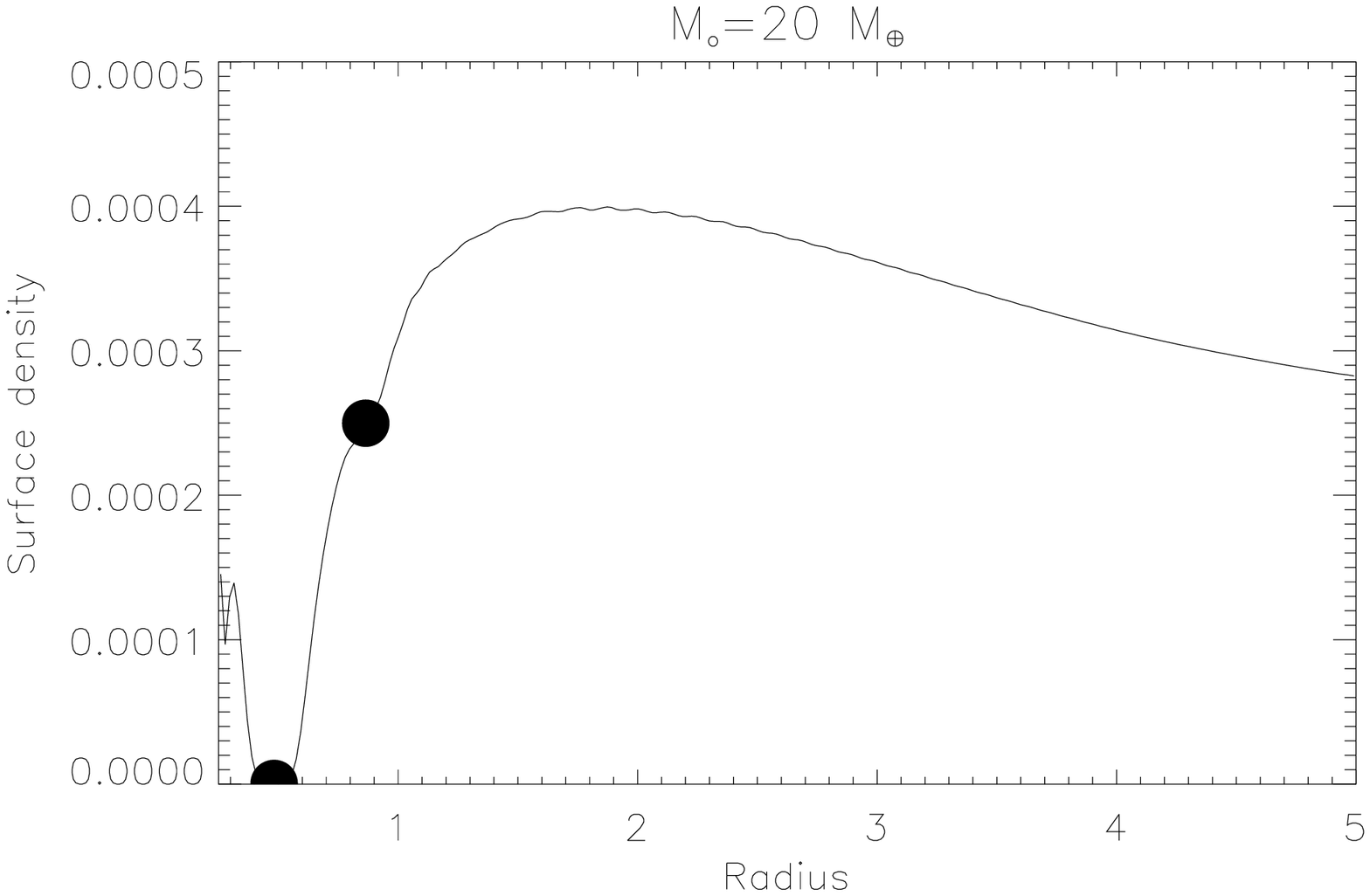}
      \caption{This figure shows the surface density profile for 
      runs $R1$ ({\it left panel}) and $R2$ ({\it right panel}) at
      $t\sim 10^4$. In
      this figure, the planets are represented by black circles.}
         \label{sigma10}
      \end{center}
   \end{figure*}

\section{Results}
 For each value of $m_o$ we consider, we have performed a simulation in which
the evolution of the system was followed for a run time of $\sim 10^4$
orbits at $r=1$. The results of the simulations are shown in Table
\ref{table1}. These indicate that, depending on the value of $m_o$,  
the evolution of the system can take three
different paths: divergent migration, such that the inner planet
migrates inward faster than the outer planet; trapping of the outer planet 
at the edge of the gap opened by the giant planet;
capture of the outer planet into a mean motion resonance with the
inner giant. We describe
in detail these three different modes of evolution in the following sections.

\subsection{Divergent migration}
Although we have not explicitly simulated the evolution of
an inner giant planet with an outer planet whose mass
is less than $10 \; \mearth$, we can see from Fig.~\ref{r1r2} 
that the typical specific torque experienced by the giant planet,
once gap formation is complete, is $\sim 9 \times 10^{-6}$.
Early on in the evolution of the $10 \; \mearth$ planet, we see
that the total specific torque is $\sim 1.8 \times 10^{-5}$.  Given
that the migration rate of a planet is related to the specific torque $T_p$ by:
\begin{equation}
\frac{dr_p}{dt}=2\sqrt{\frac{GM_\star}{r_p}}T_p,
\end{equation}
and the initial location of the Jovian mass planet is $r_p=1$ whereas 
that of the  $10\; \mearth$ body is $r_p=2.2$, we estimate that a planet of
mass $\sim 3.5\; \mearth$ will migrate at the same rate as the giant.
This suggests that an outer planet of mass $\lesssim 3.5 \; \mearth$ will
not be able to catch up with the migrating giant, and the
long term evolution will involve divergent migration.

\subsection{Trapping at the edge of the gap}

\begin{figure*}
   \begin{center}
   \includegraphics[width=0.75\columnwidth]{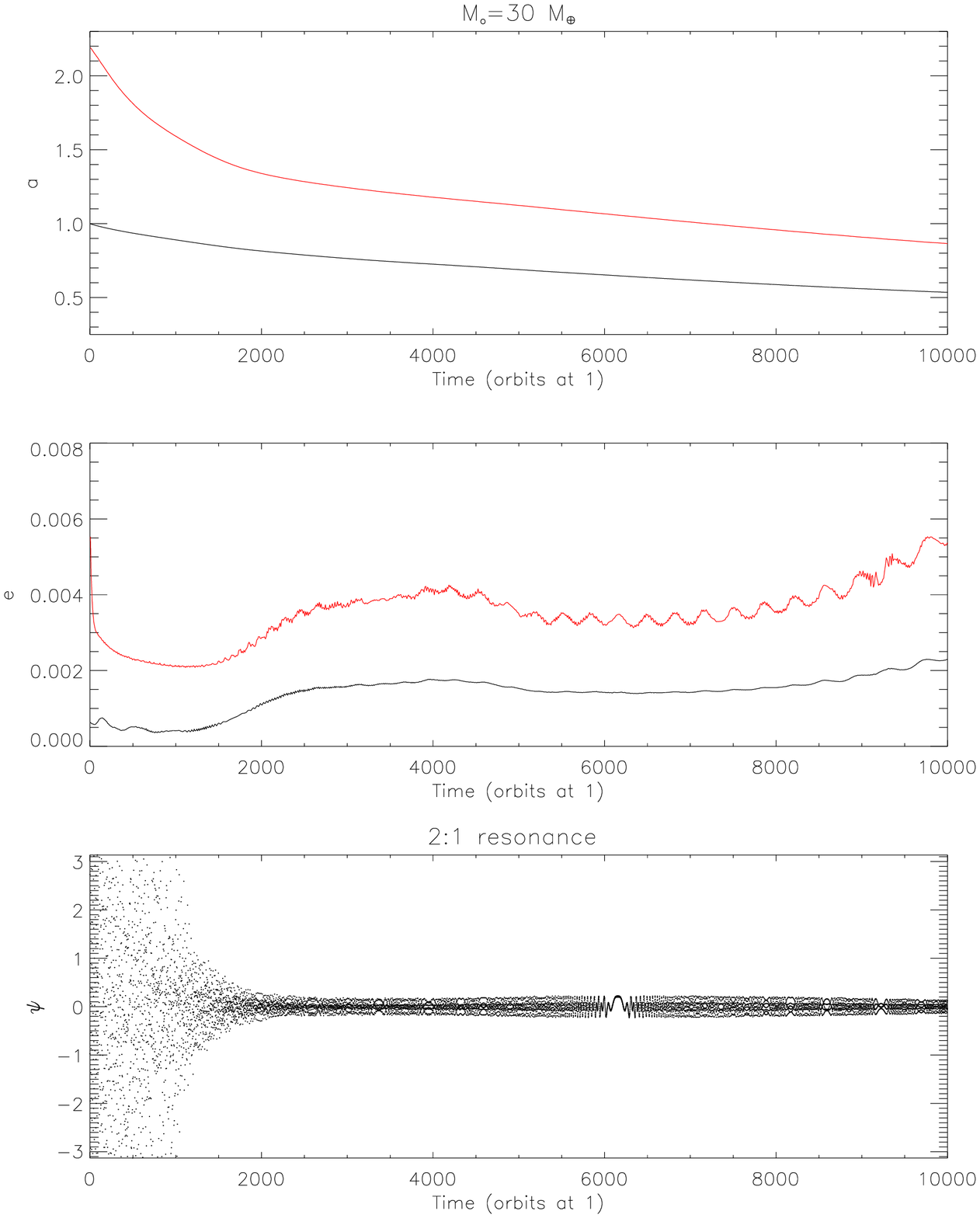}
    \includegraphics[width=0.75\columnwidth]{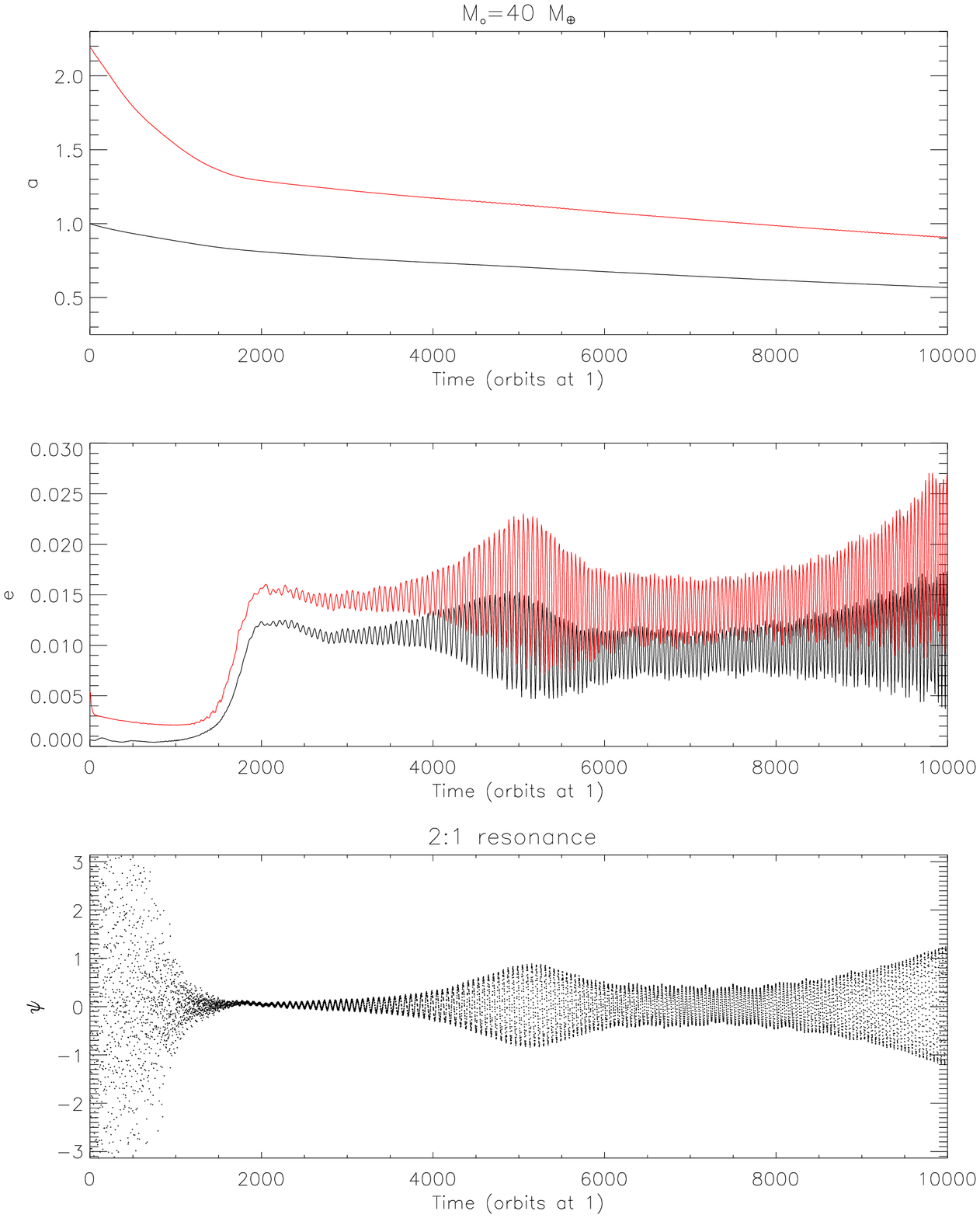}
      \caption{This figure shows the evolution of the system for
      models $R3$ ({\it left panel}) and $R4$ ({\it right panel}). In
      both cases, the giant is represented by black line and the outer
      planet is represented by red line. {\it Top:}
      evolution of the semi-major axes for both planets. {\it Middle:}
      evolution of the planets' eccentricities. {\it Bottom:}
      evolution of the resonant angle
      $\psi=2\lambda_o-\lambda_J-\varpi_J$ associated with the 2:1 resonance.}
         \label{r3r4}
     \end{center}
   \end{figure*}

\begin{figure}
   \begin{center}
   \includegraphics[width=\columnwidth,clip=true]{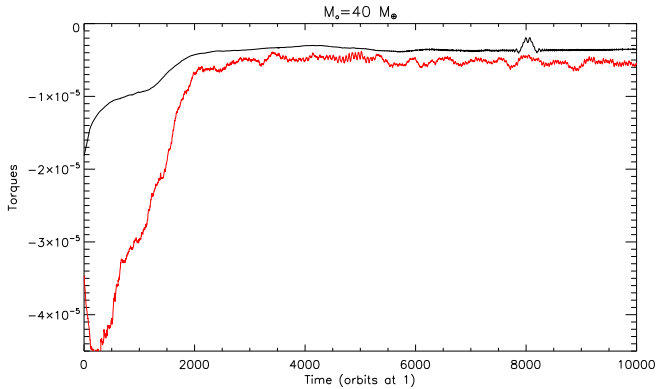}
      \caption{This figure shows, for model $R4$, the evolution of the
      disc torques exerted on the giant planet (black line) and on the
       $40\;\mearth$ planet (red line). }
         \label{torques40}
     \end{center}
   \end{figure} 

Simulations in which $m_o=10\;\mearth$ (model R1) or $m_o=20\;\mearth$
(model R2) resulted
in the outer planet being trapped at the outer edge of the giant's 
tidally maintained gap. The time evolution of 
the planets' semi-major axes for both models is shown in
the upper panel of Fig. \ref{r1r2}. In these runs, the evolution of
the system proceeds as follows. At the beginning of the
simulation, each body migrates inward as a result of disc
torques. However, as illustrated by the lower panels of 
Fig. \ref{r1r2} which display the evolution of the
specific torques acting on the planets, the type I
migration of the outer planet tends to be faster than the type II
migration of the giant. The
two planets converge as a result of this differential migration,
as illustrated by a decrease in
their period ratio  $p=(a_o/a_J)^{1.5}$, whose evolution is
depicted in the middle panel of Fig. \ref{r1r2}.
We see that after an initial period of decrease, the period ratio 
then increases for both models. This occurs from 
$t\sim 2\times 10^3$ orbits for run
$R1$, and from $t\sim 4\times 10^3$ orbits for 
simulation $R2$. Subsequently, the period ratio decreases again
and is likely to enter a sequence of cyclic variations as the
outer planets oscillate around the fixed points where
their corotation and Lindblad torques counterbalance.
This phenomenon is triggered once the outermost planet 
reaches the edge of the gap opened by the inner giant, and
suggests that such planets are unable to approach any closer to
the giant, but instead remain trapped near the gap edge. 

For model $R1$ (model $R2$), the disc surface density profile
after $\sim 10^4$ orbits  is represented on the
left (right) panel of Fig. \ref{sigma10}. Clearly, the outer
planet is located in a region of strong positive surface density
gradient, where the corotation torque is positive and can eventually
couterbalance the negative differential Lindblad torque (Masset et
al. 2006). In comparison with the 10 $\mearth$
planet, we see that the migration of the 
$20\;\mearth$ body is halted closer to
the giant, in a region  where the disc specific vorticity 
(vortensity) gradient is larger.  This arises because the specific
corotation torque scales linearly with planet mass for low mass bodies
($m_p \le 10\; \mearth$) whereas it scales with $m_p^{1/3}$ for $m_p>
50\; \mearth$. For planets in between these masses, the exponent lies
in the range $1-1/3$ (Masset et al. 2006). Therefore, a $20\; \mearth$
planet requires a steeper vortensity  gradient than a $10\; \mearth$ body to counterbalance the
Lindblad torques.\\
Interestingly, examination of the disc torques which are
represented in the  lower panel
of Fig. \ref{r1r2} reveals that once the outer planet is trapped, the 
total specific torque exerted on the latter oscillates around a mean value
corresponding to the specific torque acting on the giant.
This indicates that the outermost planet tends to follow the evolution
of the gap edge due to the migration of the giant. This
phenomenon is in good agreement with
the simulations performed by Masset et al. (2006) which showed
a 15 $\mearth$ planet remaining trapped at the edge of an expanding cavity. \\
It is well known that corotation torques can saturate in the absence
of viscosity or some other dissipative mechanism that is able to
maintain the surface density gradient in the vicinity of the planet
(Ogilvie \& Lubow 2003; Masset et al. 2006). Here, the simulations
cover a run time which greatly exceeds the horseshoe libration
time scale, and show no evidence for corotation torque saturation. In this
paper, the viscous parameter is $\alpha=10^{-3}$, which is
large enough to prevent the corotation torque saturation for planet
masses between 10-20 $\mearth$ (Masset et
al. 2006).

\begin{figure*}
   \centering
   \includegraphics[width=0.75\columnwidth]{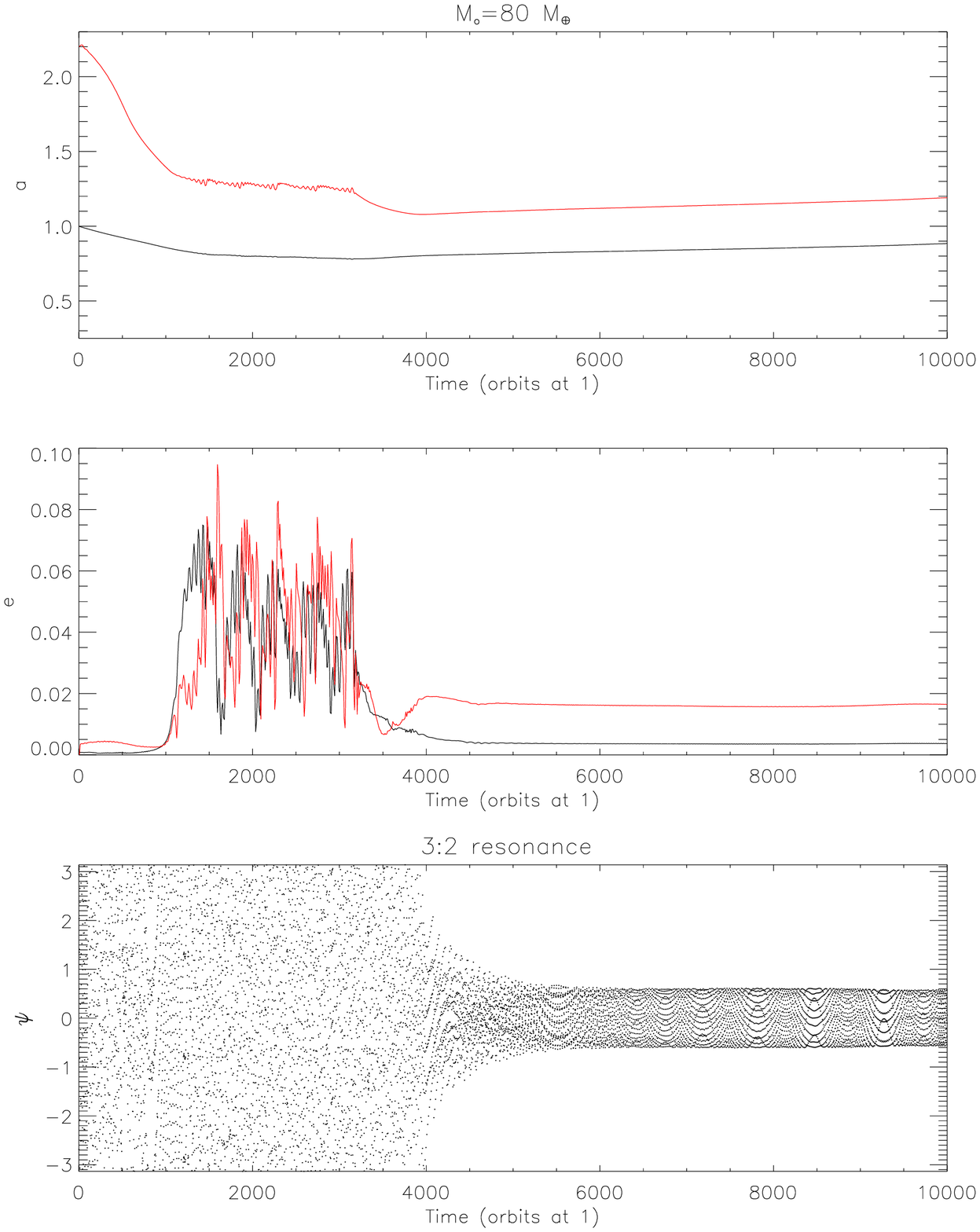}
    \includegraphics[width=0.75\columnwidth]{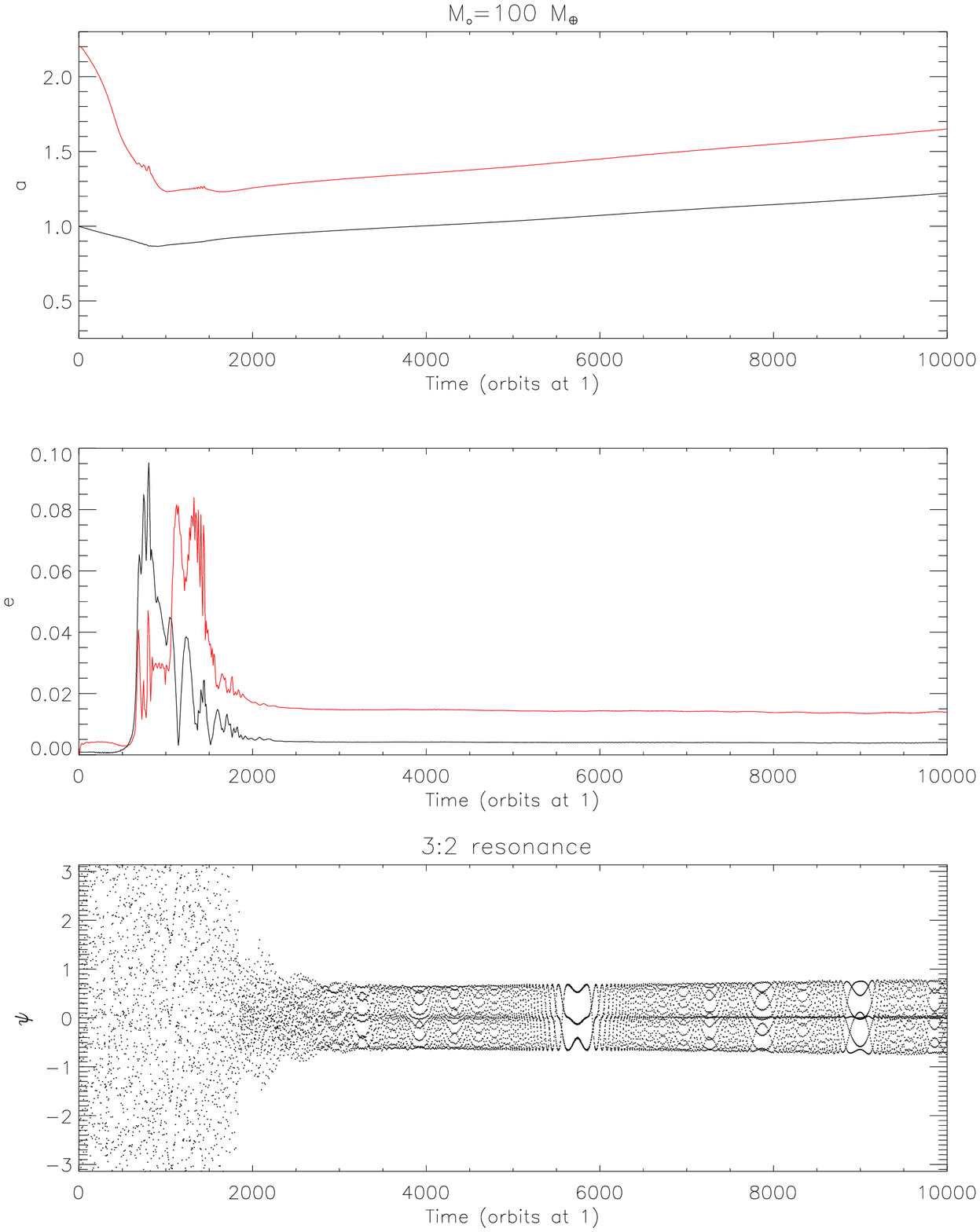}
      \caption{This figure shows the evolution of the system for
      models $R5$ ({\it left panel}) and $R6$ ({\it right panel}). In
      both cases, the giant is represented by black line and the outer
      planet is represented by red line. {\it Top:}
      evolution of the semi-major axes for both planets. {\it Middle:}
      evolution of the planets' eccentricities. {\it Bottom:}
      evolution of the resonant angle
      $\psi=3\lambda_o-2\lambda_J-\varpi_J$ associated with the 3:2 resonance.}
         \label{r5r6}
   \end{figure*}

\subsection{Capture in mean motion resonance}
\subsubsection{Models R3 and R4}

For calculations with $m_o \ge 30\;\mearth$, we find that the outer
planet can migrate through the gap region and stable mean motion
resonances are formed between the inner giant and the outermost body. The
two upper panels in Fig. \ref{r3r4} show the evolution of the
semi-major axes and eccentricities of the
planets for models $R3$ and $R4$.  In model $R3$, the mass of the outer planet
is $m_o=30\;\mearth$  while it is $m_o=40\;\mearth$ in model
$R4$. Here again,
the outer planet migrates faster initially and catches up with the 
inner body. In both cases, we find that the convergence of orbits
subsequently leads to the formation of a 2:1 resonance between the
two planets. Resonant capture occurs at $t\sim 2\times 10^3$ orbits for 
model $R3$, and at the earlier time of
$\sim 1.5\times 10^3$ orbits for
model $R4$. The lower panel of Fig. \ref{r3r4} displays the time
evolution of the resonant angle  $\psi=2\lambda_o-\lambda_J-\varpi_J$
associated with the 2:1 resonance, where $\lambda_J$
($\lambda_o$) and $\varpi_J$ are respectively the mean longitude  and
longitude of pericentre of the innermost (outermost) planet. We
see that $\psi$  undergoes libration with a smaller amplitude in 
run $R3$, which indicates that the planets are locked deeper in the
resonance in that case. Subsequent to this resonant capture, 
the continued evolution of the system in both models 
is such that the two planets migrate
inward together while maintaining the commensurability.\\
Fig. \ref{torques40} displays, for model $R4$, the evolution of
  the disc torques exerted on both the Jovian mass planet and the
  $40\; \mearth$ body. Comparing this figure with Fig. \ref{r1r2}, we
  see that for model $R4$, the torques exerted on the outermost body differ
  significantly to the ones exerted on the giant planet. This indicates that two planets locked in
  resonance exchange angular momentum in order for them to migrate at
  the same rate, which is not the case of the planets in models $R1$
  and $R2$.

\subsubsection{Models R5 and R6}

\begin{figure}
   \centering
   \includegraphics[width=0.85\columnwidth]{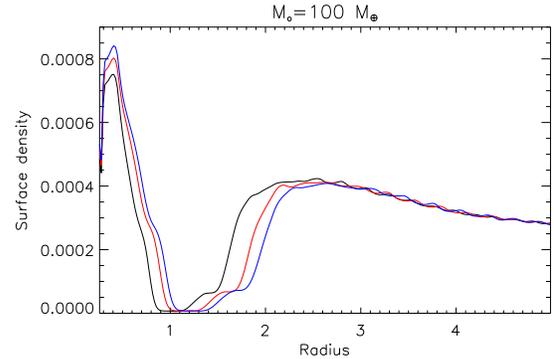}
      \caption{This figure shows the disc surface density profile for
      model $R6$ at $t=4\times 10^3$ (black line),  $8\times 10^3$
      (red line) and  $10^4$ (blue line) orbits.}
         \label{sigmar6}
   \end{figure}

The evolution of the semi-major axes and eccentricities 
of the planets for models $R5$
and $R6$ are displayed in the two upper panels of 
Fig. \ref{r5r6}. The mass of the outer planet in
model $R5$ is $m_o=80\;\mearth$ whereas
it is $m_o=100\;\mearth$ in model $R6$, 
which is characteristic of Saturn's mass. 
In both models, the outermost planet initially migrates 
in very rapidly as it
undergoes runaway migration (Masset \& Papaloizou 2003).
In run $R5$, the 80 $\mearth$ planet reaches the location of the 
2:1 resonance with
Jupiter at $t\sim 10^3$ orbits where it remains temporarily captured,
before slipping through into the 3:2 resonance
at $t\sim 4\times 10^4$ orbits. In run $R6$, the evolution of the
Saturn-mass planet is similar, although it passes
though the 2:1 resonance without being trapped. 
The evolution of the resonant
angle  $\psi=3\lambda_o-2\lambda_J-\varpi_J$ associated with the 3:2
resonance is presented, for both models, in the lower panel of
Fig. \ref{r5r6}. \\
In agreement with Masset \& Snellgrove (2001) 
who were the first to examine the
dynamics of a Jupiter plus Saturn system embedded in a
protoplanetary disc, we find that the evolution outcome 
is such that the two planets migrate outward
together maintaining the 3:2 resonance. Outward migration is favoured
because the planets tend to share a common gap as they approach each other. 
Since Lindblad torques scale as $\sim m_p^2$, the positive torques 
exerted by the inner disc on Jupiter are
larger than the negative torques acting on the
outermost planet, thereby resulting  in the outward migration
of the system (Masset \& Snellgrove 2001). Such a  process can be
maintained because there is a permanent flow of disc material from the
outer disc across the gap. This can be seen in Fig. \ref{sigmar6}
which displays the disc surface density profile for model $R6$ at
different times. Clearly, the inner disc surface density increases as
the planets migrate outward. As noted by Masset \& Snellgrove
(2001), the mass flow across the gap has two important 
effects which enable the outward
migration to be sustained. First, it continuously supplies the
inner disc with gas, thereby maintaining a large value for the inner disc
torques. Second, the gas flowing through the gap
exerts positive corotation torques on the planets, although
Morbidelli \& Crida (2007) have recently shown that their 
consequence for the
evolution of the system is sub-dominant compared with the positive
inner disc torques. \\
It is worth noting that the reversed migration of the planets may 
depend on both the
numerical and physical parameters involved. Morbidelli \& Crida (2007)
recently performed simulations of the same system and examined how
the  evolution outcome  depends on  initial conditions and
disc parameters.  They have shown that although outward migration is
robust with respect to numerical parameters, it is favoured in low
viscosity and cool discs. In the simulations presented here, we
adopted $\alpha=10^{-3}$ which is clearly low enough for the
outward migration to be maintained.\\

\subsubsection{Model R7}

\begin{figure}
   \centering
   \includegraphics[width=0.75\columnwidth]{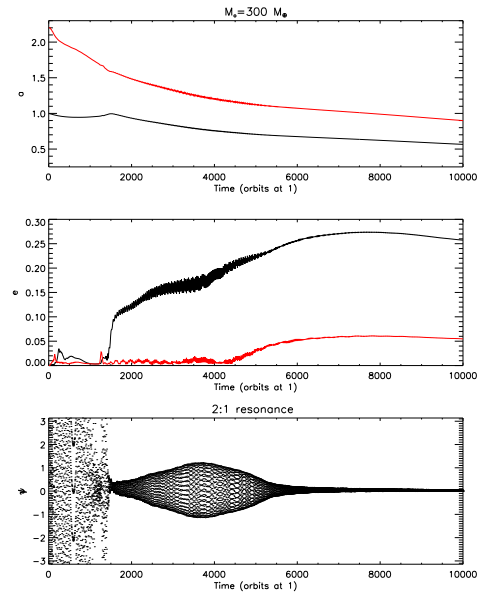}
      \caption{This figure shows the evolution of the system for
      model $R8$. {\it Top:}
      evolution of the semi-major axes for both planets. {\it Middle:}
      evolution of the planets' eccentricities. {\it Bottom:}
      evolution of the resonant angle
      $\psi=2\lambda_o-\lambda_J-\varpi_J$ associated with the 2:1 resonance.}
         \label{r7}
   \end{figure}

\begin{figure}
   \centering
   \includegraphics[width=0.75\columnwidth]{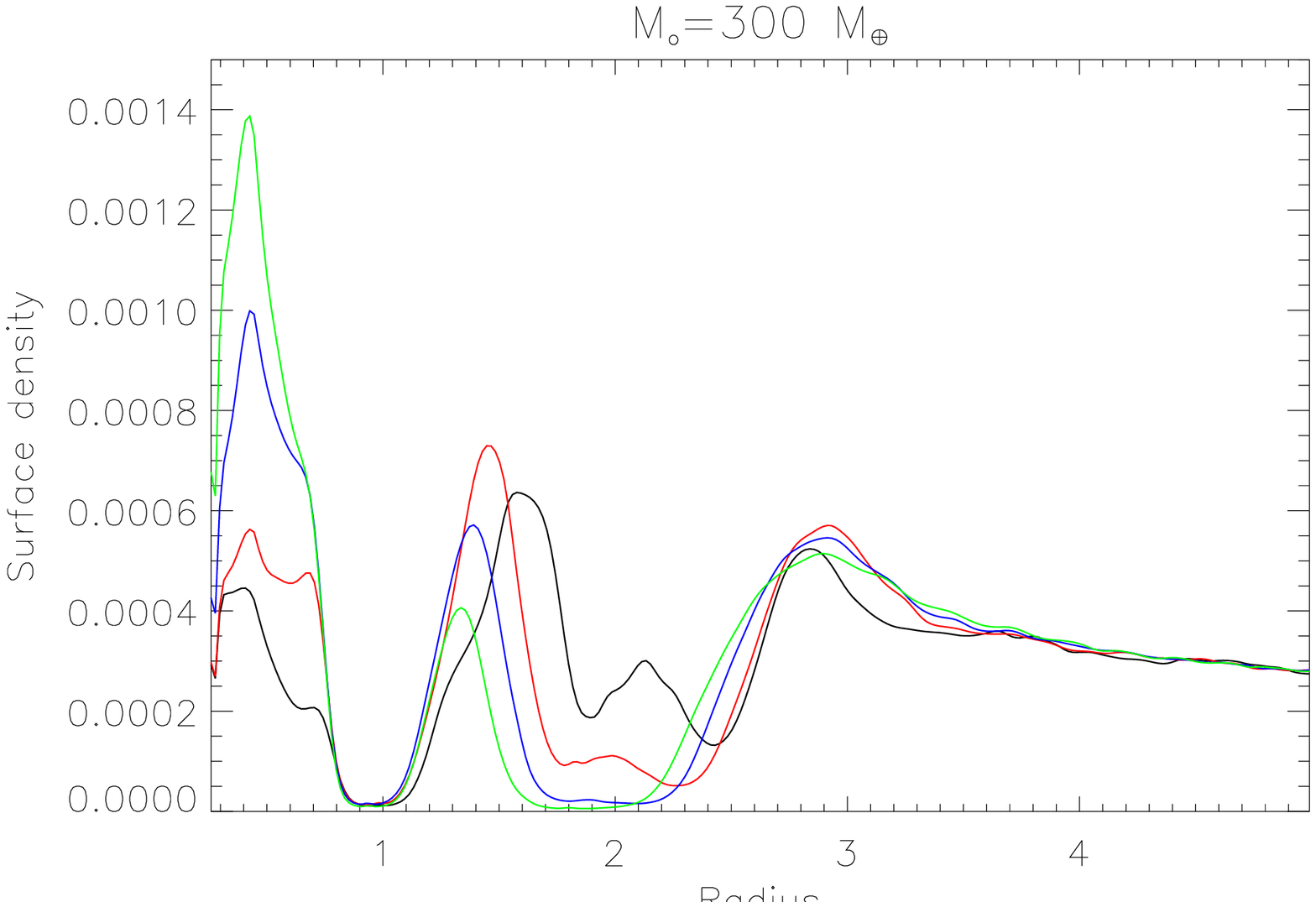}
   \includegraphics[width=0.75\columnwidth]{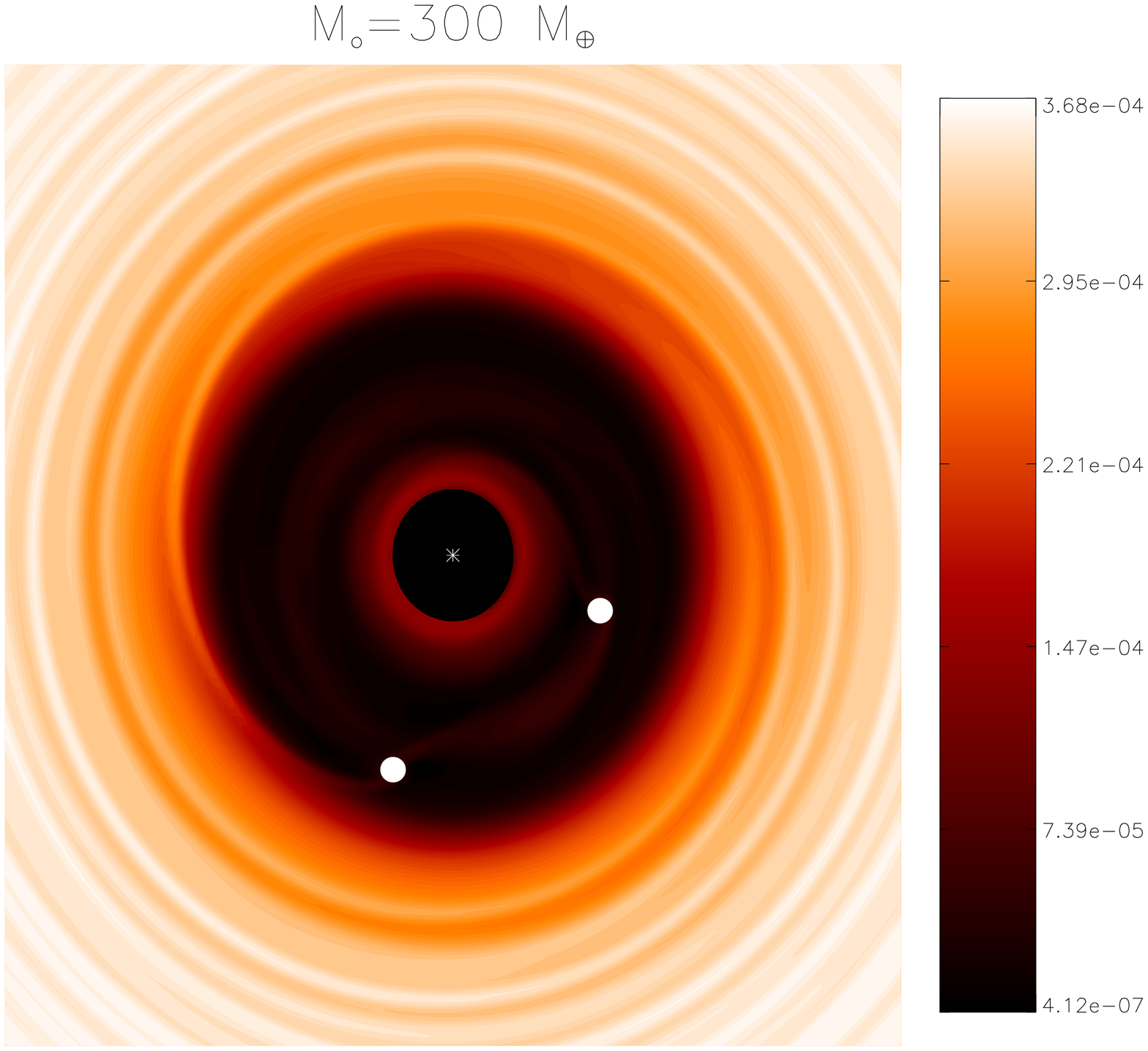}
      \caption{{\it Upper panel:} this figure shows the
      disc surface density profile for model $R7$ at $t=100$ (black
      line), 300 (red line), 600 (blue line) and 900 (green line)
      orbits. {\it Lower panel:} this figure shows in linear scale a snapshot 
      of the disc surface density
      at $t\sim 10^4$ orbits for the same model. 
      In this figure, planets are represented by white circles.}
         \label{sigmar7}
   \end{figure}

The two upper panels of Fig. \ref{r7} show the evolution of the
semimajor axes  and eccentricities of the planets for model
$R7$. Here, $m_o=1\;M_J$ so that both the inner and outermost
planets undergo slow type II migration initially. At $t\sim
800$ orbits the migration of the innermost planet is reversed,
leading to convergence of the planet orbits.
This occurs because the gas located between the two planets
tends to be cleared as the gaps formed by them join together.
The inner planet is then
positively torqued by the inner disc, but experiences a reduced
negative torque from the outer disc, causing its direction of
migration to reverse. This effect is compounded by the fact that
some of the gas originally located between the planets flows
through into the inner disc, augmenting the positive torque
it exerts. This can be
seen clearly in the upper panel of Fig. \ref{sigmar7} which 
displays the disc surface density profile at different times from the
beginning of the two--planet evolution.
The convergent migration causes the two planets to become
locked in the 2:1 resonance at $t\sim 1.5\times 10^3$ orbits. 
The time evolution of the resonant
angle  $\psi=2\lambda_o-\lambda_J-\varpi_J$ associated with the 2:1
resonance is presented in the lower panel of Fig. \ref{r7}. 
Capture into resonance  makes the eccentricities of both
planets grow rapidly before they saturate at values of
$e_J\sim 0.06$ and $e_o\sim 0.27$. Here again, the long-term
evolution of the
system appears to be inward migration with the two planets
maintaining their commensurability.  Moreover, this two--planet system tends
ultimately to evolve to a state in which the two planets
orbit within a common gap, as can be seen
in the lower panel of Fig. \ref{sigmar7} which shows a snapshot of the
disc surface density at $t\sim 10^4$ orbits.\\
The results of this calculation are broadly consistent with previous
hydrodynamical simulations of two Jovian--mass planets interacting
with their protoplanetary disc (Snellgrove et al. 2001; Papaloizou 
2003; Kley et al. 2005). However, the initial conditions that we employ
here are different from the ones used in most of the previous studies,
in which the two--planet system is assumed to initially orbit inside a
cavity. This indicates that trapping into 2:1 resonance is a robust 
outcome of such a system. 

\section{Consequences for the early Jupiter--Saturn system}
In this section we discuss the results of our simulations in
the context of recently proposed models for the early evolution of
the outer Solar System. The so--called `Nice model' (Tsiganis et al. 2005;
Morbidelli et al. 2005; Gomes et al. 2005) proposes a scenario
in which the outer planets were in a more compact configuration
just after dissipation of the gas disc, with Saturn orbiting 
interior to the 2:1 mean motion
resonance with Jupiter. Subsequent scattering of the exterior
planetesimal disc causes 
the three outer planets to migrate outward, 
during which time
Jupiter and Saturn cross their
mutual 2:1 resonance. This triggers a global instability leading to
the orbital structure of the outer Solar System observed today.
More recent work by
Morbidelli et al. (2007) shows that a similar outcome
may be obtained if Jupiter and Saturn were originally in the 3:2
resonance, with a global instability being triggered when they
cross the 5:3 resonance. They have shown that a model may be 
constructed in which all four outer Solar System planets were in
mutual mean motion resonances, and for which the net disc torques
acting on the planets is zero resulting in no migration of 
the system.

The simulations presented in this paper raise a number of questions
within the context of this scenario. The most obvious one is
whether it is possible for Saturn to have been in the 3:2 resonance
with Jupiter prior to the onset of global instability, 
if it had formed further out in the disc. Our results,
in agreement with those of Masset \& Snellgrove (2001) and
Morbidelli \& Crida (2007), show that a fully formed Saturn
is able to migrate into the 3:2 resonance. But they also show that
a lower mass core can either be trapped at the edge of Jupiter's
gap, or else be captured in the 2:1 resonance instead.
It is not clear that this core, when it grows in mass, will actually
migrate through and be locked in 3:2. Moreover, capture in the
2:1 resonance may depend on the disc density and examining whether or
not Saturn passes through the 2:1 resonance in a less massive disc
remains an outstanding issue. To investigate these
questions, we have performed an additional suite of eight simulations.

\begin{figure}
   \centering
   \includegraphics[width=\columnwidth]{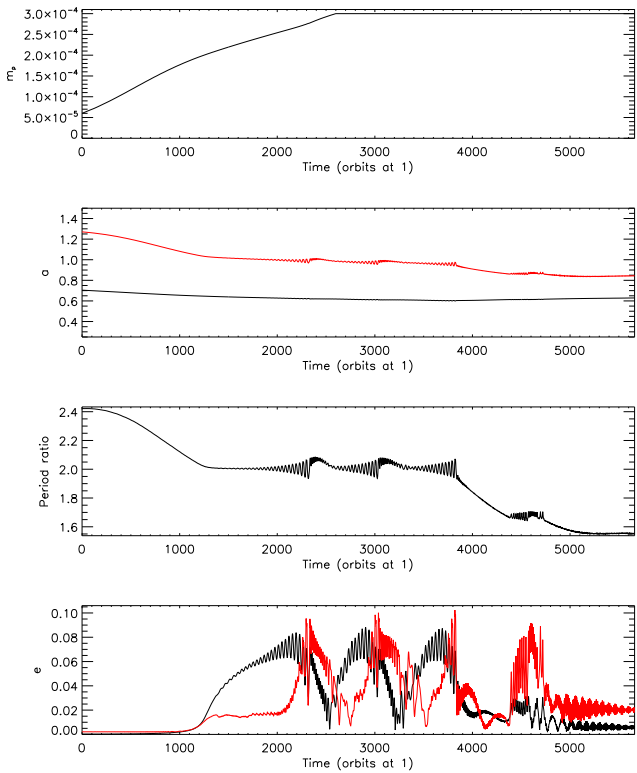}
      \caption{Evolution of a 20 $\mearth$ core
      initially trapped at the edge of Jupiter'gap and accreting gas
      material from the disc. {\it Upper (first) panel:} this figure
      shows the mass of the core as a function of time. 
      {\it Second panel:} this figure
      shows the evolution of the semi-major axes of the planets. {\it
      Third panel:} evolution of the period ratio
      $p=(a_o/a_J)^{3/2}$. {\it Lower panel:} evolution of the
      eccentricities of the planets.}
         \label{accret}
   \end{figure}

In four of these simulations we initiated the system with
a 20 $\mearth$ core trapped at the edge of Jupiter's gap.
In each simulation the core is able to accrete gas from
the disc. For each of the four runs we varied the accretion
rate  so that the time scale for the core to reach one
Saturn mass ranged between 500 to 2500 initial Jovian orbital periods.
Interestingly, each simulation resulted in the same final
outcome, with Jupiter and Saturn locked in the 3:2
resonance. Fig. \ref{accret} shows the results of a simulation in
which the core grows to a Saturn mass in $\sim 2500$
orbits. Typically, the evolution of the core proceeds as follows. At
the beginning of the simulation, the core leaves the edge of
Jupiter's gap since as it grows, the coorbital region is depleted and
corotation torques weaken. Subsequent evolution
involves further growth of the core and convergent migration of the
two planets until capture into 2:1 resonance. In each simulation
however, the accreting core breaks free from the 2:1 resonance 
once its mass has reached one Saturn mass. Continuation of
the runs indicate that the 5:3 resonance can be temporary
established but that ultimately, Saturn becomes locked in the 3:2 resonance
with Jupiter.\\
A fifth simulation was performed in which a 30 $\mearth$ core
was initially located in the 2:1 resonance with Jupiter.
Gas accretion was initiated onto the core such that it would
reach one Saturn mass in 2500 Jovian orbits. Once again the
2:1 resonance was broken as the core approached one Saturn mass,
with the planet finally settling into the 3:2 resonance.\\
Finally, we performed three additional simulations that study
  the evolution of a fully formed Saturn as a function of the disc surface
  density. In these three simulations, the initial surface density
  profile is the same as that described in Sect. \ref{sec:init}, but the
  total disc mass is reduced by factors of $2$, $4$ and $8$,
 corresponding to $M_d= 10^{-2}$,
$5\times 10^{-3}$ and $2.5\times 10^{-3} M_{\odot}$, respectively.
Reduction by a factor of $1/2$ leads to the Saturn--mass planet
passing through the 2:1 resonance, and capture in 3:2.
Reduction of the disc mass by factors of $1/4$ or $1/8$, however,
leads to a new mode
of evolution. Here, the Saturn--mass planet is captured into
the 2:1 resonance for approximately $3 \times 10^3$ orbits when
$M_d = 5 \times 10^{-3} M_{\odot}$, and for approximately $1 \times 10^4$
orbits when $M_d = 2.5 \times 10^{-3} M_{\odot}$. The resonance
then breaks in each case, and the planet is scattered outward by a small
distance such that it orbits just outside of the 2:1 resonance.
This configuration is maintained over runs times of $6 \times 10^3$
orbits, without any sign of recapture into the resonance, or
migration through it. Analysis of the disc torques experienced
by the Saturn mass planet show that they are essentially zero, apparently
because a small orbital eccentricity ($e \simeq 0.08$) is maintained,
leading to torque cancellation as the Saturn--mass planet orbits
at the edge of the gap formed by the inner giant.\\
The series of simulations presented in this section
suggest that trapping into the 3:2 resonance
is a very robust evolutionary outcome of a Jupiter--Saturn system embedded in
a protoplanetary disc, provided the disc is sufficiently massive.
In this case capture of the planet into 3:2 is found 
to be independent of the earlier
evolution of proto-Saturn. Indeed these results show that such
models cannot be used to constrain in detail the formation history
of Saturn prior to its supposed trapping in the 3:2 resonance with
Jupiter. Three possible Saturn formation scenarios exist, and
none of these can be excluded by the simulations presented in this paper: 
({\it i}) full formation of Saturn {\it in situ} close to the 3:2 resonance 
followed by capture into it; ({\it ii}) full formation of Saturn out beyond
the 2:1 resonance with Jupiter, followed by subsequent migration into
the 3:2 resonance; ({\it iii}) formation of Saturn's core and its
trapping at the edge of Jupiter's gap (or in the 2:1
resonance with Jupiter), with further growth through gas accretion causing
inward migration into the 3:2 resonance. 
The simulations performed with
significantly lower disc masses, however, indicate that capture into
3:2 would have been much less probable if Jupiter and Saturn had
formed late in the
life time of the solar nebula when the disc was being dispersed.

\section{Conclusion}
We have presented the results of hydrodynamic
simulations of two planets embedded in a protoplanetary disc.
The inner planet is a gas giant with a mass of 
$1\; M_J$, and the outer
planet has a mass that ranges between $10 \;\mearth$ to $1 \; M_J$. 
Our simulations suggest three possible evolutionary pathways for
such a system of two planets. The first involves divergent migration,
which arises when the inner planet migrates inward at a faster rate
than the outer planet. We estimate that this occurs when the 
outer planet has a mass
$m_o \lesssim 3.5 \; \mearth$. The second involves the outer planet
becoming trapped at the edge of the gap formed by the interior giant,
due to corotation torques, and arises for outer planet masses
in the range $10 \le m_o \le 20 \; \mearth$.
The third involves resonant capture of the outer planet
as it migrates toward the giant. For planet masses in the range
$30 \le m_o \le 40 \; \mearth$, and for $m_o=1 \; M_J$, capture is
into the 2:1 resonance. For planet masses in the range $80 \le m_o \le 100 \;
\mearth$, capture into the 3:2 resonance occurs. These results
are relevant to observations of extrasolar multiplanet systems,
since they suggest that there exists a lower mass limit
for the capture of an outer planet into a mean motion resonance
with an interior Jovian mass body. They suggest that Neptune--mass
planets and below should not be observed in resonance with an
interior giant planet. Planets with mass in excess of 30 $\mearth$
may, however, be observed in such resonances.

We discussed the results of our simulations within the context of
the early history of the outer Solar System, and presented some
calculations of Saturn's core being initially trapped at the edge of
Jupiter's gap (and in the 2:1 resonance) 
and growing through the accretion of gas from the disc. We demonstrated that
the final state of the Jupiter-Saturn system is likely to be trapping
in 3:2 resonance, and that such an outcome does not
depend on the earlier evolution of Saturn's core. These results
provide support for the idea that the outer planets were in a much
more compact configuration during the early evolution of the Solar
System, with Jupiter and Saturn being locked in the 3:2 resonance
(Morbidelli et al. 2007).

\begin{acknowledgements}
The simulations performed in this paper were performed on
the QMUL High Performance Computing facility purchased
under the SRIF initiative, and on the U.K. Astrophysical Fluids Facility.
\end{acknowledgements}

\end{document}